\documentclass[onecolumn,amsmath,amssymb,prd,reprint,longbibliography,floatfix,8pt]{revtex4-1}

\usepackage{cancel}
\usepackage{enumitem}
\usepackage[utf8x]{inputenc}
\usepackage{graphicx}
\usepackage{dcolumn}
\usepackage{bm}
\usepackage[switch]{lineno}
\usepackage{float} 
\usepackage{subfigure}
\usepackage{tikz}
\usepackage{multirow}
\usetikzlibrary{arrows.meta}
\usetikzlibrary{arrows,decorations.pathmorphing,backgrounds,positioning,fit,petri}

\begin{document}

\title{The role of electronic excited states in the spin-lattice relaxation of spin-1/2 molecules}

\author{Lorenzo A. Mariano$^{1}$}
\author{Vu Ha Anh Nguyen$^{1}$}
\author{Jonatan B. Petersen$^{2}$}
\author{Magnus Bj\"ornsson$^{3}$}
\author{Jesper Bendix$^{3}$}
\author{Gareth R. Eaton$^{4}$}
\author{Sandra S. Eaton$^{4}$}
\email{sandra.eaton@du.edu}
\author{Alessandro Lunghi$^{1}$}
\email{lunghia@tcd.ie}

\affiliation{$^{1}$School of Physics, AMBER and CRANN Institute, Trinity College, Dublin 2, Ireland}
\affiliation{$^{2}$Department of Chemistry, The University of Manchester, Manchester M13 9PL, United Kingdom}
\affiliation{$^{3}$Department of Chemistry, University of Copenhagen, DK-2100 Copenhagen, Denmark}
\affiliation{$^{4}$Department of Chemistry and Biochemistry, University of Denver, Denver, CO 80210, USA}

\begin{abstract}
{\bf Magnetic resonance is a prime method for the study of chemical and biological structures and their dynamical processes. The interpretation of these experiments relies on considering the spin of electrons as the sole relevant degree of freedom. By applying ab inito open quantum systems theory to the full electronic wavefunction, here we show that contrary to this widespread framework the thermalization of the unpaired electron spin of two Cr(V) coordination compounds is driven by virtual transitions to excited states with energy higher than 20,000 cm$^{-1}$ instead of solely involving low-energy spin interactions such as Zeeman and hyperfine ones. Moreover, we found that a window of low-energy THz phonons contributes to thermalization, rather than a small number of high-energy vibrations. This work provides a drastic reinterpretation of relaxation in spin-1/2 systems and its chemical control strategies, and ultimately exemplifies the urgency of further advancing an ab initio approach to relaxometry.}
\end{abstract}

\maketitle

\section*{Introduction}

The ubiquitous presence of spin in chemical compounds, either because of the presence of spin-active nuclei or unpaired electrons, makes magnetic resonance a central experimental technique for the study of chemical and biological processes. In particular, the time it takes for a spin to reach its thermal equilibrium state, namely the spin relaxation time, provides unique insights into the interactions between the spin and its chemical environment. For instance, the study of spin relaxation time provides insights into the dynamics of the molecules they are embedded in, which can be exploited to determine the aggregation state of bio-molecules\cite{eaton2000relaxation}, their folding dynamics\cite{parigi2021unveiling}, the presence of analytes in solutions\cite{li2019all}, and enhance in-vivo imaging techniques\cite{alzola2022comprehensive}. In recent years, relaxometry experiments have also become a central tool of quantum sensing\cite{degen2017quantum}, where the relaxation of highly coherent spin states is used to probe magnetic fields and magnetic states with unprecedented sensitivity\cite{mzyk2022relaxometry}.

\begin{figure*}
    \begin{center}
      \begin{tikzpicture}
        \node[anchor=south west, inner sep=0] (image) at (0,0) {\includegraphics[width=0.8\textwidth]{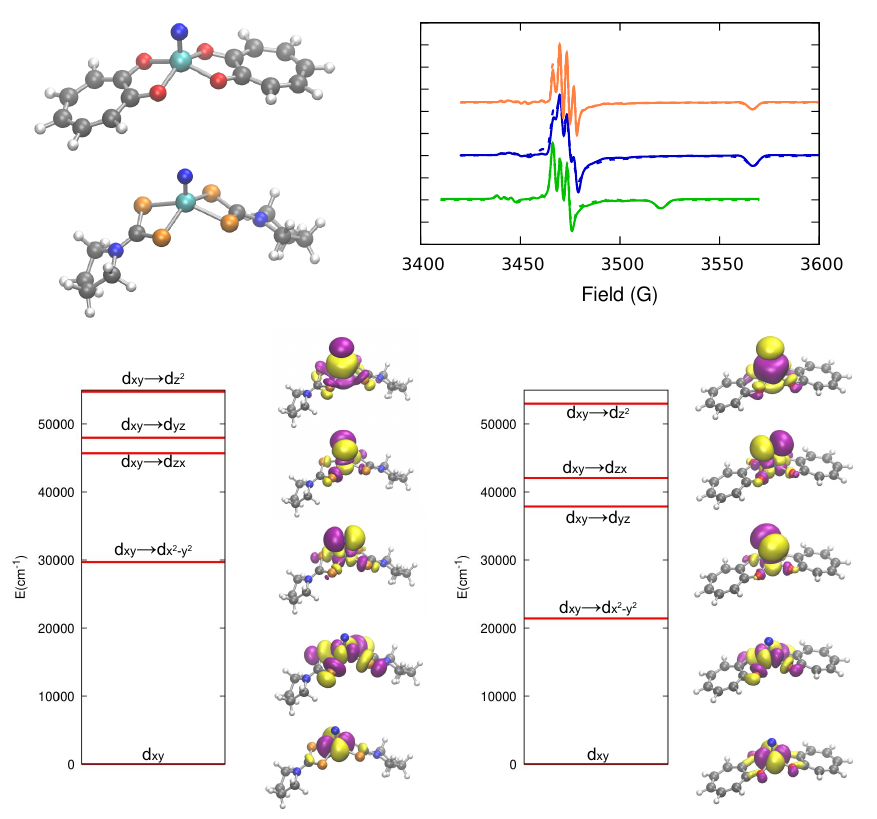}};
        \node at (0.5,13) {\large{$\mathbf{A}$}};
        \node at (6.2,13) {\large{$\mathbf{C}$}};
        \node at (0.5,7.5) {\large{$\mathbf{B}$}};
      \end{tikzpicture}
    \end{center}  
    \caption{\textbf{(A) Molecular structures.} CrN(trop)$_2$ (top) and CrN(pyrdtc)$_2$ (bottom). Color code: cyan for Cr, blue for N, grey for C, red for O, orange for S, white for H.
    \textbf{(B) Absorption spectra.} Simulated absorption peaks for CrN(pyrdtc)$_2$ (left) and CrN(trop)$_2$ (right). CASSCF molecular orbitals for each state are also reported using an isovalue of $0.02$ $e$/\AA$^3$. Positive (negative) values of the wavefunction are plotted in purple (yellow). \textbf{(C) EPR spectra.} Simulated (solid line) and measured (dashed line) cw-EPR spectum for CrN(pyrdtc)$_2$ in green, CrN(trop)$_2$ in blue, and CrN(trop-d4)$_2$ in orange.  }
    \label{absCr}
\end{figure*}
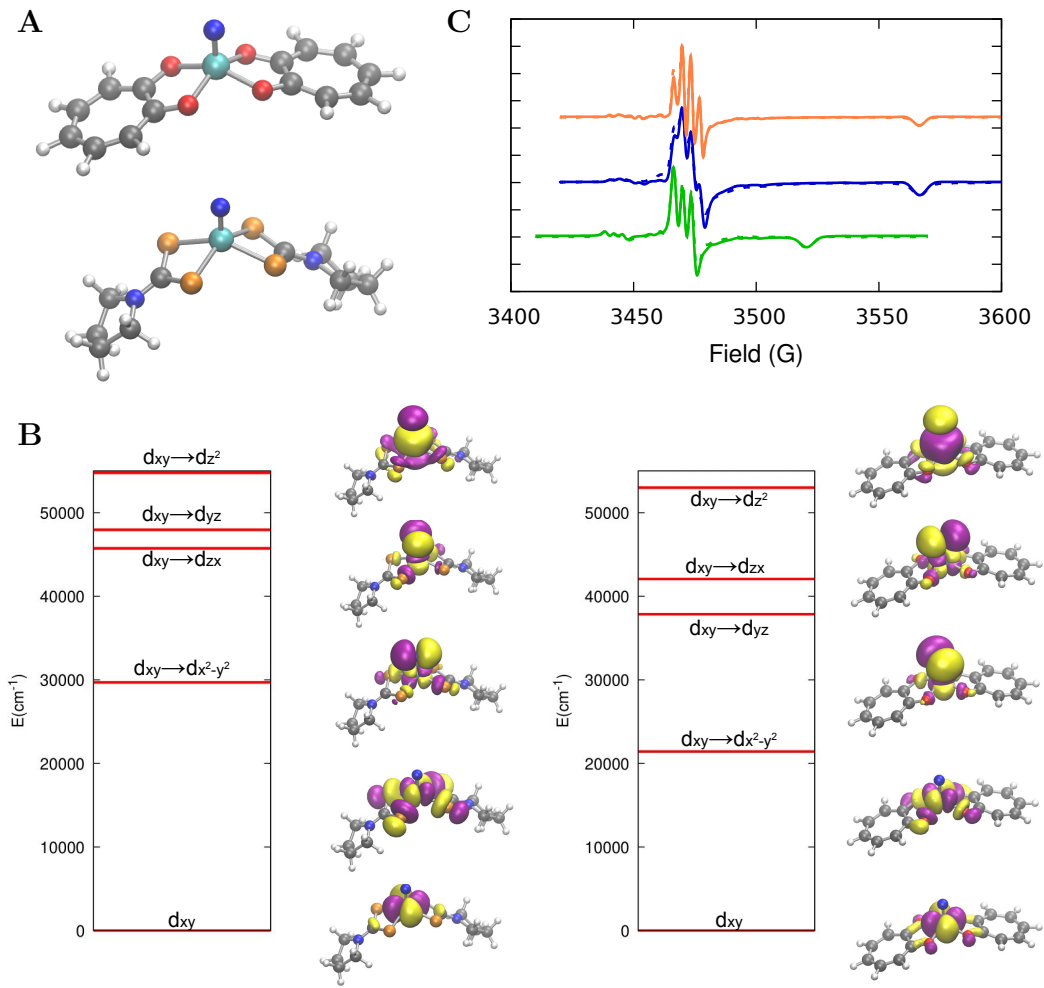

However, the information that can be extracted from these experiments is only as good as the theoretical framework used to interpret experimental evidence. The canonical quantum theory of spin relaxation is based on a perturbative treatment of how the thermal motion of atoms inside a crystalline lattice, i.e. the phonons, affects the crystal field of unpaired electrons located in the 3$d$ or 4$f$ shell of an ion\cite{van1940paramagnetic}. Following Orbach's work in the '60s\cite{orbach1961spin}, this complex quantum mechanical framework is generally reformulated in terms of a phenomenological parametric description of the sole ground-state multiplet, namely the $2S+1$ possible orientations associated with a spin of magnitude $S$. This effective ground-state approach to spin-lattice relaxation theory still represents the cornerstone of the interpretation of experiments.

Despite its success, such a phenomenological theory of spin-lattice relaxation can only partly assist the interpretation of relaxometry experiments due to the fact that it is not a predictive tool and does not make it possible to unequivocally determine the leading relaxation mechanism, nor the interaction at the origin of relaxation itself. Only recently, these striking limitations have been overcome by combining spin-lattice relaxation theory with ab initio methods\cite{lunghi2023spin}, where all the coefficients appearing in the equations describing the dynamics of spin are fully determined from first principles. Such an ab initio theory of spin relaxation has already provided a successful prediction of both one- and two-phonon spin relaxation rates for a diverse range of systems, going from coordination compounds of 3$d$ and 4$f$ ions with long relaxation times\cite{lunghi2022toward,mondal2022unraveling,reta2021ab,nabi2023accurate}, known as single-molecule magnets, to solid-state quantum sensors based on color centers, such as defects in diamond and hexagonal boron-nitride\cite{mondal2023spin,cambria2023temperature}. A common denominator for all these compounds is a spin moment larger than 1/2 and energy levels defined by zero-field splitting.

Despite the momentous success of ab initio spin dynamics, the description of spin relaxation in two-level spin-1/2 compounds\cite{lunghi2019phonons,lunghi2020limit}, arguably the most fundamental type of system, has proved to be the toughest challenge, and fundamental questions about its nature remain unanswered. In particular, none of the fundamental interactions involving a typical spin-1/2 system seems to be able to fully explain its dynamics at temperatures above $\sim 20$ K (Raman relaxation), with the Zeeman and the hyperfine interactions failing to explain i) the time-scale of relaxation and the absence of a correlation with the external magnetic field intensity\cite{lunghi2020limit,lunghi2022toward}, and ii) the angular dependence of relaxation rates\cite{kazmierczak2022illuminating} and their correlation with spin-orbit coupling strength\cite{chakarawet2021effect}, respectively. In the attempt to reconcile some of these aspects of theory and experiments, the involvement of electronic excited states has been invoked\cite{kazmierczak2022illuminating}, but a full ab initio assessment of their role to Raman relaxation and a comparison with other proposed relaxation mechanisms are still missing. Overall, such a state of affairs casts serious doubts on our understanding of this fundamental physical process and must be urgently addressed.

Here, we show that this conundrum can be solved by lifting the fundamental assumption of Orbach’s canonical approach to spin-lattice relaxation theory, namely by going beyond considering the sole spin-1/2 molecular ground state and by considering virtual transitions to high-energy electronic excited. We show that this novel theoretical framework can reproduce the relaxation times of two prototypical spin-1/2 Cr(V) coordination compounds as measured by electron paramagnetic resonance (EPR) inversion recovery experiments, advancing the establishment of a complete ab initio theory of spin relaxation. Moreover, we review recent results on other spin-1/2 coordination compounds and critically discuss the need to include all possible mechanisms in a comprehensive approach to spin-lattice relaxation theory.

\section*{Results}

The compounds nitrido bis(pyrrolidine dithiocarbamate)chromium(V) (\textbf{CrN(pyrdtc)$_2$}) and nitrido bis(tropolone)chromium(V) (\textbf{CrN(trop)$_2$}) are chosen for our investigation of the role of the various contributions to Raman relaxation in spin-1/2. In this work we adopt a very inclusive definition of Raman relaxation as any process that simultaneously involves two phonons\cite{lunghi2023spin}. Among the many molecules with spin-1/2, Cr(V) systems are particularly advantageous for our study because samples with natural nuclear isotopic abundance are comprised of 90.5\% $^{52}$Cr with nuclear spin moment $I = 0$ and 9.5\% $^{53}$Cr with $I = 3/2$. This permits the evaluation of the impact of nuclear hyperfine interaction on relaxation in the same sample by selectively measuring one or the other nuclear isotope.  CrN(trop)$_2$ and its deuterated counterpart also allow the study of the effect of a rigid auxiliary coordination sphere and nuclear spin-free ligands on spin relaxation time, which have been flagged as possible sources of decoherence\cite{zadrozny2015millisecond,lunghi2020limit}.\\

\textbf{Synthesis, electronic structure and magnetism.} The synthesis follows the general protocol developed for nitrido-chromium(V) complexes, which involves atom-transfer (formally transfer of N$^-$) from N,N'-ethylene bis(salicylideniminato)nitrido manganese(V), MnN(salen), to semi-labile CrCl$_3$(THF)$_3$ in poorly coordinating solvents. Good yields over 70\% are obtained consistently for synthesis carried out in dry acetonitrile. The synthesis of CrN(pyrdtc)$_2$ was first reported in ref \cite{birk2003atom}, whilst CrN(trop)$_2$ is new among tropolonate complexes. The solid-state structural parameters determined from single crystal X-ray diffraction follow similar trends among nitrido complexes. The molecular structures are depicted in Fig. \ref{absCr}A. The complex with deuterated tropolonate (CrN(trop-d4)$_2$), is also synthesized following the same procedure. Full deuteration of all positions except the most distant 5-position in the tropolone ring is achieved as previously reported for other derivatives\cite{sakakibara2021direct}. Finally, the compound ReN(pyrdtc)$_2$, isomorphous to the chromium complex, is synthesized\cite{DT9720001079} to serve as a diamagnetic solid-state host for CrN(pyrdtc)$_2$. Dilution of paramagnetic species into a diamagnetic host removes dipolar spin-spin interactions that interfere with the measurement of spin-lattice relaxation.\\

The electronic excitations for CrN(pyrdtc)$_2$ and CrN(trop)$_2$ are computed with multireference methods and reported in Fig. \ref{absCr}B together with their description in terms of molecular orbitals, revealing a single unpaired electron located in the $d_{xy}$ orbital of their Cr(V) ion. Excited states corresponding to intra-band $d$ transition are present at high energy, starting from $\sim$ 30,000 cm$^{-1}$ for CrN(pyrdtc)$_2$ and $\sim$ 20,000 cm$^{-1}$ for CrN(trop)$_2$, revealing a well-isolated spin-1/2 Kramers doublet ground state. Simulations are consistent with spectroscopic studies\cite{birk2003atom,bendix1998nitridocyanometalates}, apart from an overall overestimation of the transitions' energies. 

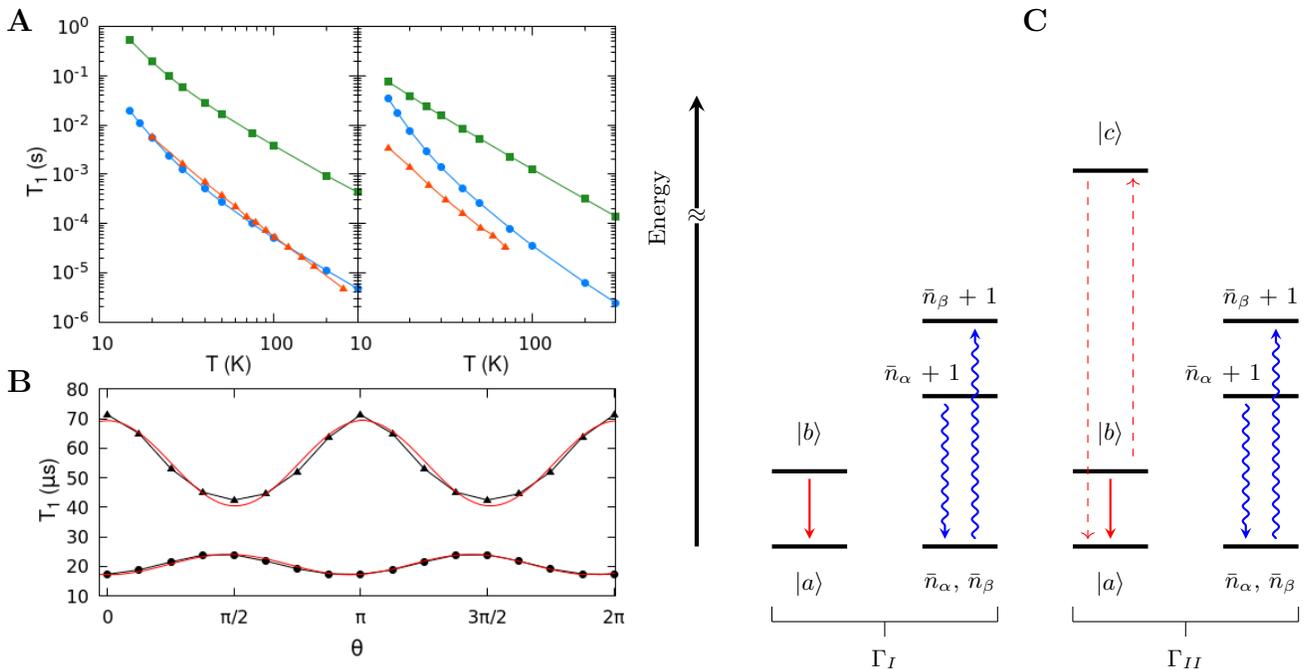
\begin{figure*}
 
\begin{tikzpicture}


\node[anchor=south west, inner sep=0] (image) at (-4,0.5) {\includegraphics[width=0.5\textwidth]{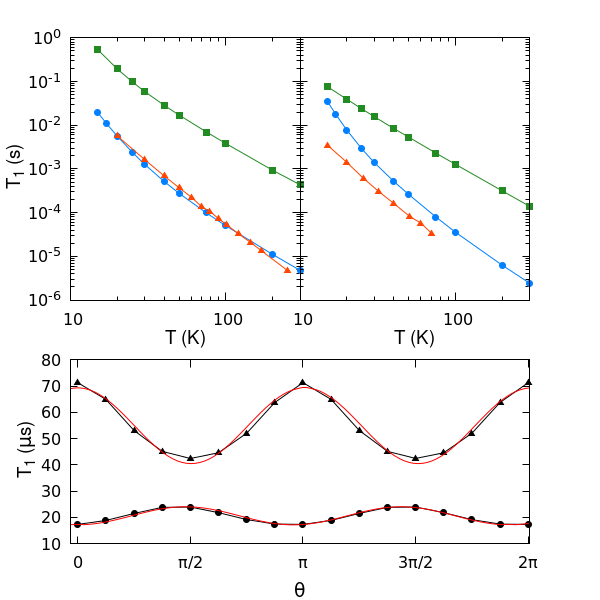}};


\node at (-4,9) {\large{$\mathbf{A}$}};
\node at (-4,4.2) {\large{$\mathbf{B}$}};
\node at (9.5,9) {\large{$\mathbf{C}$}};


\def\s{7}
\def\sr{5}


\draw [ultra thick] (-2+\s,2) -- (-2+\s,6.3);
\node at (-2+\s,6.43) {$\sim$};
\node at (-2+\s,6.35) {$\sim$};
\draw [ultra thick, ->,>=stealth] (-2+\s,6.5) -- (-2+\s,8);
\node [rotate=90] at (-2.5+\s,6.5) {Energy};


\draw (-1+\s,1) -- (2+\s,1);
\draw (2+\s,1) -- (2+\s,1.2);
\draw (-1+\s,1) -- (-1+\s,1.2);
\draw (0.5+\s,0.8) -- (0.5+\s,1);
\node at (0.5+\s,0.5) {$\Gamma_I$};

\draw [ultra thick] (-1+\s,2) -- (0+\s,2);
\draw [ultra thick] (-1+\s,3) -- (0+\s,3);

\draw [ultra thick] (1+\s,2) -- (2+\s,2);
\draw [ultra thick] (1+\s,4) -- (2+\s,4);
\draw [ultra thick] (1+\s,5) -- (2+\s,5);

\draw [thick,red,<-,>=stealth] (-0.5+\s,2.1) -- (-0.5+\s,2.9) ;  
\draw [thick,blue,->,>=stealth,decorate, decoration={snake,amplitude=.4mm,segment length=2mm,post length=1mm}] (1.3+\s,3.9) -- (1.3+\s,2.1) ;  
\draw [thick,blue,->,>=stealth,decorate, decoration={snake,amplitude=.4mm,segment length=2mm,post length=1mm}] (1.7+\s,2.1) -- (1.7+\s,4.9) ;  

\node at (-0.5+\s,1.5) {$| a \rangle$};
\node at (-0.5+\s,3.5) {$| b \rangle$};

\node at (1.5+\s,1.5) {$\bar{n}_{\alpha}$, $\bar{n}_{\beta}$};
\node at (1+\s,4.3) {$\bar{n}_{\alpha}$ + 1 };
\node at (1.5+\s,5.3) {$\bar{n}_{\beta}$ + 1 };


\draw (5+\sr,1) -- (8+\sr,1);
\draw (8+\sr,1) -- (8+\sr,1.2);
\draw (5+\sr,1) -- (5+\sr,1.2);
\draw (6.5+\sr,0.8) -- (6.5+\sr,1);
\node at (6.5+\sr,0.5) {$\Gamma_{II}$};

\draw [ultra thick] (5+\sr,2) -- (6+\sr,2);
\draw [ultra thick] (5+\sr,3) -- (6+\sr,3);
\draw [ultra thick] (5+\sr,7) -- (6+\sr,7);

\draw [ultra thick] (7+\sr,2) -- (8+\sr,2);
\draw [ultra thick] (7+\sr,4) -- (8+\sr,4);
\draw [ultra thick] (7+\sr,5) -- (8+\sr,5);

\draw [thick,red,<-,>=stealth] (5.5+\sr,2.1) -- (5.5+\sr,2.9) ;

\draw [dashed,<-,red] (5.2+\sr,2.1) -- (5.2+\sr,6.9) ;
\draw [dashed,<-,red] (5.8+\sr,6.9) -- (5.8+\sr,3.2) ;

\draw [thick,blue,->,>=stealth,decorate, decoration={snake,amplitude=.4mm,segment length=2mm,post length=1mm}] (7.3+\sr,3.9) -- (7.3+\sr,2.1)  ;  
\draw [thick,blue,->,>=stealth,decorate, decoration={snake,amplitude=.4mm,segment length=2mm,post length=1mm}] (7.7+\sr,2.1) -- (7.7+\sr,4.9)  ;  

\node at (5.5+\sr,1.5) {$| a \rangle$};
\node at (5.5+\sr,3.5) {$| b \rangle$};
\node at (5.5+\sr,7.5) {$| c \rangle$};

\node at (7.5+\sr,1.5) {$\bar{n}_{\alpha}$, $\bar{n}_{\beta}$};
\node at (7+\sr,4.3) {$\bar{n}_{\alpha}$ + 1 };
\node at (7.5+\sr,5.3) {$\bar{n}_{\beta}$ + 1 };

\end{tikzpicture}
\caption{\textbf{(A) Spin-lattice relaxation time.} The relaxation times for CrN(pyrdtc)$_2$ and CrN(trop)$_2$ are reported in the left and right panels, respectively. Red lines and triangles correspond to experimental results obtained with inversion recovery at the $g_\perp$ field and X-band frequency. Blue lines and dots report simulation results obtained with $\Gamma_{II}$. Green lines and squares report simulation results obtained with $\Gamma_{I}$. \textbf{(B) $\mathbf{T_1}$ angular dependency.} The relaxation time computed at 100 K as a function of the direction of the applied magnetic field is reported. Triangles and circles correspond to CrN(pyrdtc)$_2$ and CrN(trop)$_2$, respectively. Continuous red lines show the fitting of the computed points using $f(\theta)=sin^2(\theta)$. \textbf{(C) Schematic representation of spin-phonon transitions.} The left panel describes a spin transition due to a resonant two-phonon process, where one phonon is emitted and one is simultaneously absorbed. The right panel describes a spin transition among the same, where a two-phonon process occurs thanks to virtual transitions (red dashed lines) to an excited electronic state $|c\rangle$.}
\label{tauCr}
\end{figure*}

Thanks to the large energy separation between ground and excited states, these compounds represent the prototypical systems where their magnetic properties can be very accurately described by an effective spin-1/2 Hamiltonian, and all excited states are assumed to play negligible roles up to room temperature. The X-band continuous wave (CW) spectra of CrN(pyrdtc)$_2$, CrN(trop)$_2$ and CrN(trop-d4)$_2$, reported in Fig. \ref{absCr}C, were interpreted with a canonical spin Hamiltonian
\begin{equation}
    \hat{H}_\mathrm{s}=\mu_\mathrm{B} \: \vec{\mathbf{S}} \cdot  \mathbf{g} \cdot \vec{\mathbf{B}} + \vec{\mathbf{S}} \cdot  \mathbf{A} \cdot \vec{\mathbf{I}}\:,
    \label{spinH}
\end{equation}
where the electron spin operator $\vec{\mathbf{S}}$ is coupled to the external magnetic field $\vec{\mathbf{B}}$ and the $^{53}
$Cr's nuclear spin $\vec{\mathbf{I}}$ through the Bohr's magneton, $\mu_\mathrm{B}$, the Lande' tensor, $\mathbf{g}$, and the hyperfine coupling tensor, $\mathbf{A}$, respectively. The dominant features are from the $I = 0$ $^{52}$Cr isotope which is about 90.5\% of natural abundance. Weaker, but well-defined lines are observed for the $I = 3/2$ $^{53}$Cr isotope. The eigenvalues of $\mathbf{g}$ and $\mathbf{A}$ obtained by simulating the EPR spectra are summarized in Table 1 and exhibit characteristic axial anisotropy. Such symmetry axis roughly coincides with the CrN chemical bond and will be referred to as the $g_\parallel$ direction. The $g_\perp$ values for CrN(pyrdtc)$_2$ and CrN(trop)$_2$ are similar, but the $g_{\parallel}$ value for CrN(trop)$_2$ is significantly lower than for CrN(pyrdtc)$_2$. 
The comparison with values for other Cr(V) nitrido complexes that have been reported previously in the literature\cite{konda1994electron,du1995electron} shows substantial similarities. The eigenvalues of $\mathbf{g}$ and $\mathbf{A}$ for CrN(pyrdtc)$_2$ in 3:7 CH$_2$Cl$_2$:toluene and doped into the Re analog are the same within experimental uncertainty, demonstrating that the two different environments do not significantly change the molecular structure. 

\begin{table}[h!]
\begin{tabular}{l c|c|c|c}
 &  & X-Band & DFT & NEVPT2 \\ \hline
 \multirow{6}{*}{CrN(pyrdtc)$_2$} & $g_{iso}$ & 1.9875 & 1.984 & 1.983 \\ 
 & $g_{\parallel}$ & 1.9673  & 1.969  & 1.955 \\ 
 & $g_{\perp}$ & 1.9976  & 1.992  & 1.996 \\ \cline{2-5}
 & $A_{iso}$ & 74.20  & 79.31 & -- \\ 
 & $A_{\parallel}$ &  -- & 125.71 & -- \\ 
 & $A_{\perp}$ & 49.84  & 57.33 & -- \\ \hline
 \multirow{6}{*}{CrN(trop)$_2$} & $g_{iso}$ & 1.9678 & 1.976 & 1.971 \\
 & $g_{\parallel}$ & 1.9415 & 1.953 & 1.925 \\ 
 & $g_{\perp}$ & 1.9958 & 1.986 & 1.994 \\ \cline{2-5}
 & $A_{iso}$ & 74.76 & 82.61 & -- \\ 
 & $A_{\parallel}$ & -- & 128.66 & -- \\ 
 & $A_{\perp}$ & 49.84 & 62.48 & -- \\ \hline
\end{tabular}
\caption{\textbf{Spin Hamiltonian parameters.} The table reports the eigenvalues of \textbf{A} and \textbf{g} from Eq. \ref{spinH} obtained by fitting the CW-EPR spectra and simulated with DFT and NEVPT2. Hyperfine couplings are reported in MHz. Experimental values for CrN(pyrdtc)$_2$ and CrN(trop)$_2$ were obtained at 60-70 K for the compound diluted in the Re analog and in glassy toluene:CH$_2$Cl$_2$, respectively.}
\label{spinCoeff}
\end{table}

Table \ref{spinCoeff} also reports the eigenvalues of the tensors appearing in Eq. \ref{spinH} calculated by two different electronic structure theory methods. As observed in other studies, Density Functional Theory (DFT) methods are well suited to recover electronic dynamical correlation in systems with a single unpaired $d$ electrons and achieve a good agreement for both $\mathbf{g}$ and $\textbf{A}$ tensors\cite{singh2018challenges,garlatti2023critical}. NEVPT2 is found to overestimate the $g_\perp$ shifts but still correctly capture the overall experimental trends. The calculation of $\mathbf{A}$ at this level of theory has not been attempted, but recent work suggests this is possible\cite{birnoschi2024relativistic}. In conclusion, both methods adequately describe the electronic structure of these compounds. \\

\textbf{Spin-lattice relaxation measurement.} Spin-lattice relaxation for CrN(pyrdtc)$_2$ and CrN(trop)$_2$ are measured by 3-pulse inversion recovery as detailed in the Materials and Methods section. In these experiments, the direction of the electron spin is inverted with respect to the direction of the magnetic field, and the time it takes to relax back to the thermal equilibrium, namely $T_1$, is monitored. Emphasis is placed on data obtained at temperatures above about 20 K, where the Raman process is proposed to dominate\cite{eaton2000relaxation,eaton2018relaxation}. Results obtained at the X-band frequency (see Fig. \ref{tauCr}A) reveal a sharp decrease of $T_1$ as a function of temperature, in agreement with the vast literature on spin-1/2 transition metal complexes\cite{konda1994electron,bader2014room,zadrozny2015millisecond,atzori2016room}. Faster relaxation for CrN(trop)$_2$ than for CrN(pyrdtc)$_2$ is consistent with the general trend that sees compounds with larger $g$-shift (deviation from the free electron value of 2.0023) relax faster due to a larger effective spin-orbit coupling\cite{sato2007impact}. The differences in $T_1$ between the two compounds can also be partly attributed to the different host matrix, Re-doped solid and glassy solution, respectively. The latter environment likely presents softer lattice vibrations, which are known to speed up spin relaxation \cite{sato2007impact}. Values of $T_1$ for CrN(trop)$_2$ in 3:7 CH$_2$Cl$_2$:toluene are indistinguishable from CrN(trop-d4)$_2$ in 3:7 CD$_2$Cl$_2$:toluene-d$_8$ which shows that deuteration plays a negligible role.

Experiments were repeated for different values of the external static magnetic field, isolating the contribution of molecules with different orientations, different field intensities, and different nuclear isotopes. The comparison between experiments conducted at X-band and Q-band conclusively shows that the overall intensity of the static magnetic field up to $\sim 1.2$ T does not substantially affect $T_1$ for the same orientation between field and molecule. Although it is not feasible to obtain data that define the full orientation dependence of $T_1$ for randomly oriented molecules in the glassy or powdered samples, it is possible to probe molecules oriented parallel or perpendicular to the external field. For CrN(pyrdtc)$_2$ in the Re analog the ratio of $T_1$ for the parallel and perpendicular directions is 1.4 at X-band and 1.8 at Q-band, therefore pointing to longer values of $T_1$ observed when the field is along $g_\parallel$. In glassy toluene:CH$_2$Cl$_2$ the ratio is 1.2. For CrN(trop)$_2$ the ratio of $T_1$ (parallel/perpendicular) is 1.6 at X-band and 1.5 at Q-band. Finally, the comparison between $T_1$ for the two Cr's isotopes does not reveal any substantial difference, supporting the negligible role of hyperfine interactions in the spin relaxation process for these complexes in rigid matrices. \\

The experimental $T_1$ vs $T$ curves are interpreted with conventional empirical models. According to the latter, above about 10 or 20 K the dominant contributions for most spin-1/2 species are Raman processes due to a Debye-like distribution of phonons plus a local-mode phonon with a discrete energy, usually above the Debye cutoff temperature\cite{eaton2000relaxation,eaton2018relaxation}. The temperature dependence of $T_1$ for CrN(pyrdtc)$_2$ in ReN(pyrdtc)$_2$ and for CrN(trop)$_2$ in 3:7 CH$_2$Cl$_2$ could be fit equally well with two models: i) two local modes with energies of 55 and 200 cm$^{-1}$ for CrN(pyrdtc)$_2$ and 42 and 145 cm$^{-1}$ for CrN(trop)$_2$, or ii) replacement of the lower energy local modes with a Raman process due to a Debye distribution of phonons with Debye temperatures of 70 and 55 cm$^{-1}$ for CrN(pyrdtc)$_2$ and CrN(trop)$_2$, respectively. Both models indicate that a substantial range of phonon energies are required to match the data. The ability to fit the data with two rather different models of the phonon distribution is a caution that a match with the temperature dependence of the experimental data does not constitute proof of the validity of the proposed phonon energy distribution responsible for the relaxation. A comparison of the temperature dependence of $T_1$ for CrN(pyrdtc)$_2$ and CrN(trop)$_2$ with other Cr(V) shows substantial similarities in the temperature dependence of $T_1$ across compounds and lattices suggest that the calculations and interpretations in this paper have important implications for other Cr(V) complexes.\\

\textbf{Spin-lattice relaxation theory and simulations.} According to the canonical picture of spin-lattice relaxation the interactions appearing in the spin Hamiltonian of Eq. \ref{spinH} are modulated by the activity of phonons, $q_{\alpha}$\cite{orbach1961spin}. Historically, this physical model was based on the ideal picture of extended ionic crystal lattices described by a Debye distribution. However, ab initio simulations\cite{lunghi2019phonons,lunghi2020multiple} and experiments\cite{moseley2018spin,garlatti2020unveiling,chiesa2020understanding,garlatti2023critical} have recently made it possible to describe the real distribution of phonons in molecular crystals, showing large deviations from this ideal picture. In particular, very low-energy optical phonons are present at energies well below the common Debye temperature of ionic compounds. The latter are characterized by a strong admixing of intra-molecular and rotational components which can efficiently modulate molecular spin-orbit interactions\cite{lunghi2019phonons}, ultimately making it possible for energy to flow between spin and lattice. This physical picture of phonons in molecular crystals also helps to understand why the relaxation rate of a molecule is not drastically affected by different environments, e.g. crystals and frozen solutions, where optical phonons retain much of the molecular normal modes of vibration and rotational motions are not very environment-specific. This coupling between spin and phonons is formally represented by the operators
\begin{equation}
    \hat{V}^{\alpha}= \left( \frac{\partial \hat{H}_\mathrm{s}}{\partial q_{\alpha}} \right) \quad \text{and} \quad \hat{V}^{\alpha \beta}=  \left( \frac{\partial^2 \hat{H}_\mathrm{s}}{\partial q_{\alpha}\partial q_{\beta}} \right) \:,
    \label{sph}
\end{equation}
where one or two phonons are modulating the spin's properties, respectively. \\

In previous work\cite{lunghi2020limit,lunghi2022toward,garlatti2023critical}, the second-order terms of Eq. \ref{sph} have been used to derive the Raman relaxation rate as a function of temperature through the expression
\begin{equation}
\Gamma_{I} = \frac{\pi}{2\hbar^{2}}  \sum_{\alpha\ge\beta} |V^{\alpha\beta}_{ab}|^2 \: G(\omega_{ab},\omega_{\alpha},\omega_{\beta}) \:, \label{Red22}
\end{equation}
where
\begin{equation}
    G(\omega_{ba},\omega_{\alpha},\omega_{\beta}) = \delta(\omega_{ba}-\omega_{\alpha}+\omega_{\beta})\bar{n}_{\alpha}(\bar{n}_{\beta}+1) \:,
    \label{G2sph}
\end{equation}
and $V^{\alpha\beta}_{ab}$ is the matrix elements of the quadratic coupling operator in Eq. \ref{sph} among the eigenstates of $\hat{H}_\mathrm{s}$. The latter states approximately correspond to the parallel and antiparallel orientation of the spin along the magnetic field direction. $E_{a}$ is the energy of said spin states, $\hbar\omega_{ab}=E_a-E_b$, $\hbar\omega_\alpha$ is the energy of the $\alpha$-phonon, and $\bar{n}=[\mathrm{exp}(\hbar\omega/\mathrm{k_{B}}T)-1]^{-1}$ is the Bose-Einstein thermal phonon population. For a two-level system, the relaxation time is connected to Eq. \ref{Red22} as $T_1 = (2\Gamma_I)^{-1}$. Fig. \ref{tauCr}C provides a graphical representation of this process. A transition between two spin states, e.g. a spin flip, is induced by two simultaneous phonon processes, where one is absorbed by the spin and the other emitted. To fulfil energy conservation, introduced by the presence of a Dirac delta in Eq. \ref{G2sph}, the energy difference between the two phonons must match the spin energy splitting. 

Eq. \ref{Red22} is used to simulate $T_1$ as a function of temperature, field intensity, and field orientation to compare with the experimental data from inversion recovery experiments. When the separated contributions of \textbf{A} and \textbf{g} to Eq. \ref{sph} are considered, the modulation of the former is predicted to be the leading relaxation mechanism at all fields investigated in this study ($\sim$ 0.33 and 1.2 T) and to lead to values of $T_1$ that are independent of the magnitude of the external magnetic field. The modulation of \textbf{g} instead leads to $T_1 \propto B^{-2}$, which is inconsistent with the observed similarity of $T_1$ values at X-band ($\sim$ 0.33 T) and Q-band (1.2 T). When studying the dependence of $T_1$ on the magnetic field orientation, we observe that both the modulation of \textbf{A} and \textbf{g} lead to orientation anisotropy. All these results are in agreement with former reports for several Vanadyl compounds at both X- and Q-band (\textit{vide infra} Fig. \ref{tauVO})\cite{lunghi2020limit,garlatti2023critical}. However, differently from previous studies, a relaxation mechanism due to the modulation of \textbf{A} in these compounds is inconsistent with the similarity in $T_1$ for the two nuclear isotopes of Cr ($I = 0$ and $I = 3/2$). Moreover, a large deviation between predictions of $T_1$ and experiments is observed for the Cr(V) compounds, with the predictions based on Eq. \ref{Red22}  overestimating the $T_1$ by two orders of magnitude (see Fig. \ref{tauCr}A). In conclusion, neither terms of Eq. \ref{spinH} is fully consistent with experimental data.\\

We shall now consider an alternative mechanism of Raman relaxation. When the linear coupling of Eq. \ref{sph} is combined with fourth-order time-dependent perturbation theory\cite{lunghi2022toward,lunghi2023spin}, Raman relaxation is described by a rate population transfer among spin states that reads
\begin{align}
 \Gamma_{II} =  \frac{\pi}{2\hbar^{2}} \sum_{\alpha \ge\beta} \Big| T^{\alpha\beta,+}_{ab} + T^{\beta\alpha,-}_{ab}  \Big|^{2} G(\omega_{ab},\omega_{\alpha},\omega_{\beta}) \:,
    \label{Red41}
\end{align}
where
\begin{equation}
    T^{\alpha\beta,\pm}_{ab}= \sum_{c}
    \frac{ V^{\alpha}_{ac} V^{\beta}_{cb} }{E_{c}-E_{b}\pm\hbar\omega_{\beta}}\:.
    \label{Red41_T}
\end{equation}
All the terms in Eqs. \ref{Red41} and \ref{Red41_T} carry the same meaning as in Eq. \ref{Red22}, with the additional presence of a sum over the excited states, $c$. The latter commonly take the name of virtual states to stress that although they correspond to real molecular electronic states, the process described by Eq. \ref{Red41} does not involve their population. This relaxation mechanism is graphically described in Fig. \ref{tauCr}C, where the same spin and phonon transition are concerned, but the process is now mediated by virtual transitions. 

Eq. \ref{Red41} has been successfully used to simulate spin relaxation in molecules with spin larger than 1/2\cite{lunghi2022toward,mondal2022unraveling}. However, for two-level systems such as the Cr(V) compounds, the use of Eq. \ref{Red41} is prevented by the absence of any excited state $c$. It is then apparent that Eq. \ref{Red41} can only contribute to the relaxation rate of a spin-1/2 system if electronic excited states beyond the spin Hamiltonian formalism are included. We therefore apply Eq. \ref{Red41} by reinterpreting the matrix elements of the first-order derivatives of Eq. \ref{sph} as
\begin{equation}
    \hat{V}^{\alpha}_{ab}=\langle \varphi_a | \left( \frac{\partial \hat{H}_\mathrm{el}}{\partial q_{\alpha}} \right) | \varphi_b \rangle\:,
    \label{spinphonH}
\end{equation}
where the state vectors $\{| \varphi_a \rangle\}$ represent the eigenstates of the full electronic Hamiltonian $\hat{H}_\mathrm{el}$, inclusive of spin-orbit coupling. These include the spin-1/2 ground-state doublet and all electronic excited states represented in Fig. \ref{absCr}B. These new matrix elements of Eq. \ref{spinphonH} are computed with a multireference ab initio method recently developed by some of the authors\cite{mariano2023manual}. 

\begin{figure*}
   \begin{center}    
    \begin{tikzpicture}
     \node[anchor=south west, inner sep=0] (image) at (0,6.5) {\includegraphics[scale=0.65]{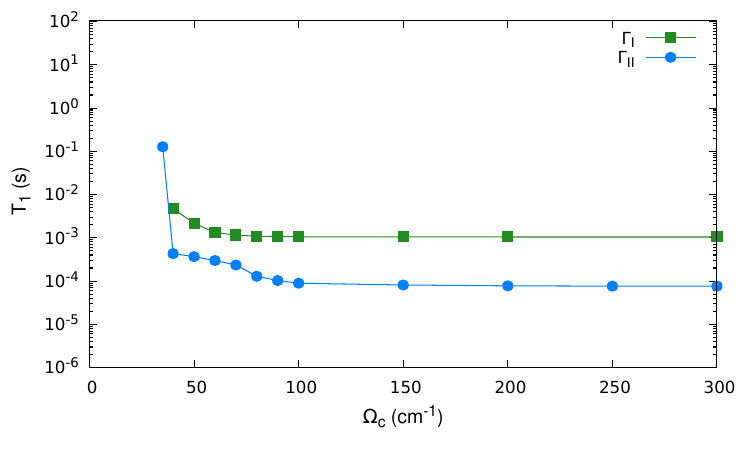}};
     \node[anchor=south west, inner sep=0] (image) at (8,6.5) {\includegraphics[scale=0.65]{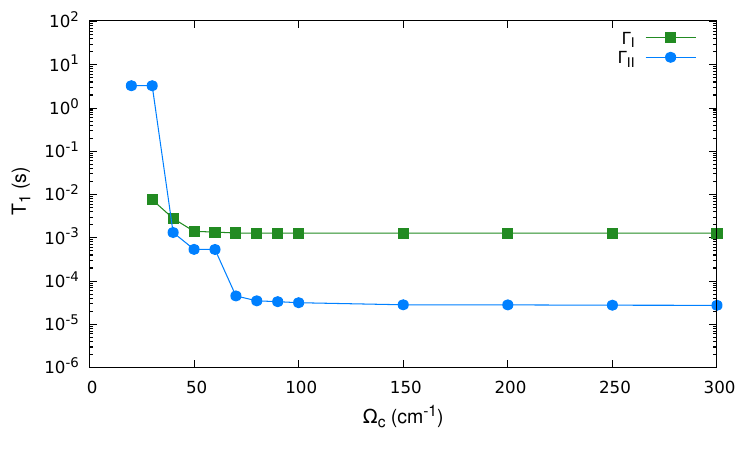}};
      \node[anchor=south west, inner sep=0] (image) at (0.3,0) {\includegraphics[scale=0.55]{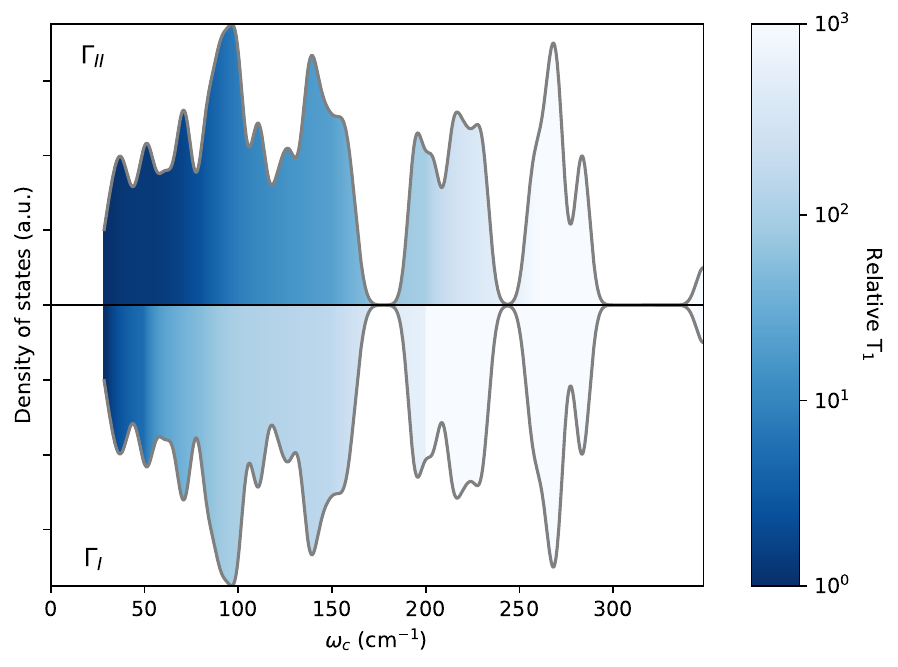}};
      \node[anchor=south west, inner sep=0] (image) at (8.7,0)  {\includegraphics[scale=0.55]{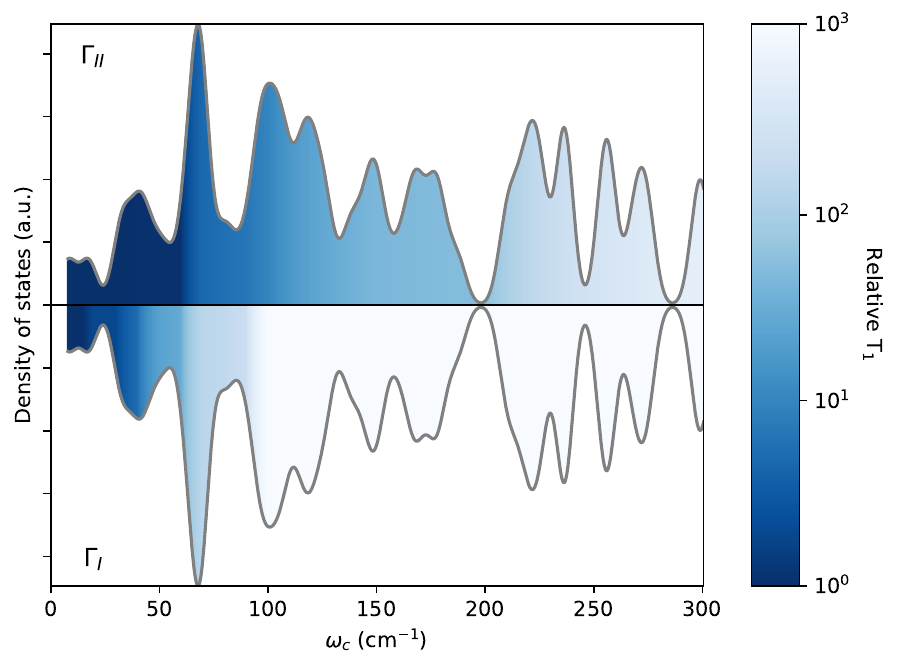}};
      \node at (0.2,11.1) {\large{$\mathbf{A}$}};
      \node at (0.2,5.9) {\large{$\mathbf{B}$}};
     \end{tikzpicture}
    \end{center} 
    \caption{\textbf{(A) High-energy phonon energy cutoff dependence of relaxation time.} The value of $T_1$ at 100 K is reported for CrN(pyrdtc)$_2$ (left) and CrN(trop)$_2$ (right). $T_1$ is computed with $\Gamma_I$ (green squares and line) and  $\Gamma_{II}$ (blue squares and line) by excluding phonons with $\omega > \Omega_c$. \textbf{(B) Low-energy phonon cutoff dependence of relaxation time.} Phonon density of states are reported for CrN(pyrdtc)$_2$ (left) and CrN(trop)$_2$ (right). The curves are shaded according to the contribution to the computed relaxation time $T_1$ at 100 K, with exclusion of phonons with $\omega < \omega_c$, where the latter corresponds to the value reported on the x-axis. Coloring of the phonon distributions above the center line and reflected below the center line correspond to the relaxation times obtained with $\Gamma_{II}$ and $\Gamma_I$, respectively, and normalised to the value obtained including all the phonons in the simulation.}
    \label{phon_convergence}
\end{figure*}

As depicted in Fig. \ref{tauCr}A, simulated values of $T_1$ using Eq. \ref{Red41} and the redefinition of Eq. \ref{spinphonH} are in excellent agreement with experimental values for both compounds, with the only notable slight difference being the slope of the curves. A slightly larger deviation between simulations and experiments is observed for CrN(trop)$_2$, which we attribute to comparing simulations of crystal phonons and measurements in frozen solution. Fig. \ref{tauCr}B  also reports the value of $T_1$ as a function of the angular dependence of the external magnetic field. We found excellent agreement between simulations and experimental results for compound CrN(pyrdtc)$_2$. In the case of CrN(trop)$_2$, we observe a reverse dependence, with longer $T_1$ for the perpendicular orientation. However, we note that the orientation dependency of $T_1$ in CrN(trop)$_2$ is very small in both simulations and experiments, making a direct comparison less straightforward. Simulations also correctly reproduce the $\sin^2(\theta)$ orientational dependency of $T_1$ recently observed in the literature for other spin-1/2 molecules\cite{kazmierczak2022illuminating,kazmierczak2023t}. Finally, this mechanism of Raman relaxation is found to be independent of the magnitude of the external magnetic field and the nuclear isotope, therefore fitting all experimental evidence. 

Finally, we address the important question of identifying the relevant phonons for spin relaxation. To determine the most relevant vibrational energy window, the simulation of $T_1$ at 100 K is performed including all phonons up to a cutoff $\Omega_{c}$. The full results of Fig. \ref{tauCr}A are recovered for $\Omega_{c} \rightarrow \infty$. Fig. \ref{phon_convergence}A reports the results of this analysis for both compounds clearly showing that predicted values of $T_1$ are converged by $\Omega_c \sim 70$ cm$^{-1}$ for both mechanisms considered ($\Gamma_I$ and $\Gamma_{II}$), meaning that phonons of higher energy do not contribute to the observed values of relaxation time. This result is in agreement with previous considerations based on calculations, where the optical phonons with the lowest energy were identified as the dominant contribution to relaxation\cite{lunghi2020limit,lunghi2022toward,garlatti2023critical}. This result is understood by noticing two key facts: i) the thermal population of phonons, $\bar{n}$, decreases rapidly as a function of their energy for a given $T$, making high-energy phonons less effective, and ii) that a large density of states is observed in the low energy part of the vibrational spectrum of the lattice (reported in Fig. \ref{phon_convergence}B). We now turn to the question of how much $T_1$ would increase if we removed the lowest energy phonons, e.g. by synthetic design. To answer this question we perform the simulation of $T_1$ at 100 K including all phonons down to a lower cutoff $\omega_{c}$. When $\omega_{c}$ is set to a lower value than the first available phonon, the full results of Fig. \ref{tauCr}A are recovered. By shifting $\omega_{c}$ to higher values, we exclude the low energy phonons from the computation and observe how $T_1$ is affected. Fig. \ref{phon_convergence}B reports the results of this analysis for both compounds. Through the color scale, we can observe that the value of $T_1$ predicted through $\Gamma_I$ increases drastically as $\omega_{c}$ is increased. When we consider the contribution of phonons through $\Gamma_{II}$ we observe a slightly different scenario, with a slower increase of $T_1$ as $\omega_{c}$ is increased, effectively pointing to the fact that even though the lowest energy phonons are still the ones that determine $T_1$ through a winner-takes-it-all effect, a large window of vibrations provides only slightly less effective alternative relaxation pathways and that phonons up to $\sim$ 120 cm$^{-1}$ would need to be removed to extend $T_1$ by one order of magnitude 100 K. Interestingly, the thermal population of phonons enters the expression of $\Gamma_I$ and $\Gamma_{II}$ in identical ways and the explanation for such a difference between the two mechanisms must be found in the different pre-factors of the function $G$. Indeed, the expression of $\Gamma_{II}$ shows that phonons at high energy are favoured by minimizing the denominators of the functions $T^{\alpha\beta,\pm}_{ab}$. Such condition is however weaker than the one imposed by the phonon thermal population in virtue of the fact that the former is a quadratic expression and the latter is exponential. As a result, phonons with energy higher than the lowest ones offer more efficient alternative relaxation pathways in $\Gamma_{II}$ than in $\Gamma_{I}$.

\section*{Discussion}

\begin{figure}[h!]
    \centering    
    \includegraphics[scale=0.72]{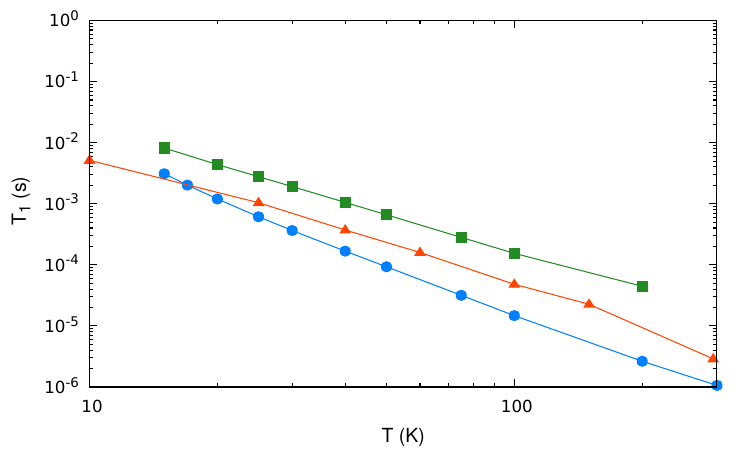} 
    \caption{\textbf{VO(dmit)$_2$ Spin-lattice relaxation time.}  Red lines and triangles correspond to experimental results obtained with inversion recovery at X-band frequency. Blue lines and dots report simulation results obtained with $\Gamma_{II}$. Green lines and squares report simulation results obtained with $\Gamma_I$.}
    \label{tauVO}
\end{figure}

Despite their conceptual simplicity, the description of spin relaxation in spin-1/2 molecules has proven to be the hardest nut to crack among a variety of magnetic systems\cite{lunghi2023spin}. Spin lattice relaxation times for a variety of organic radicals and transition metal complexes have been observed to decrease as the $g$-shift values increase\cite{sato2007impact,eaton2000relaxation, eaton2018relaxation}. Early modeling of $T_1$ and based on a spin Hamiltonian ansatz elevated this experimental correlation to causation and attempted to quantitatively predict spin relaxation in spin-1/2 in terms of the modulation of the $g$-tensor\cite{albino2019first,mirzoyan2020dynamic}. Both previous\cite{lunghi2019phonons,lunghi2020limit,lunghi2022toward} and present computational results have however shown that the modulation of the $g$-tensor cannot account for spin relaxation in spin-1/2 molecules, due to its spurious field dependence\cite{eaton2001frequency,atzori2016quantum} and leading to overall too long values of $T_1$. Without abandoning the conceptual framework of the spin Hamiltonian, the hyperfine interaction stood out as the only interaction able to explain spin relaxation in spin-1/2 systems, and despite the independence of $T_1$ on Cr's nuclear spin is not consistent with this mechanism, the modulation of \textbf{A} has managed to adequately explain all experimental evidence in the case of Vanadyl compounds  (see Fig. \ref{tauVO})\cite{lunghi2020limit,lunghi2022toward,garlatti2023critical}

As we have shown here, the solution to this puzzle can be found by overcoming the limits of Steven's effective spin Hamiltonian formalism to interpret the dynamical properties of spin systems\cite{elliott1952theory}. The pervasive use of this theoretical framework can be traced back to Orbach's seminal paper on the use of Stevenson effective operators to interpret spin-lattice relaxation in rare-earth salts\cite{orbach1961spin} and has become the fundamental language of magnetic resonance ever since.
The results obtained in this work for CrN(pyrdtc)$_2$ and CrN(trop)$_2$ demonstrate that molecular electronic excited states instead contribute to the Raman virtual transitions described by $\Gamma_{II}$, dominating two-phonon spin relaxation at a temperature above $\sim$ 20 K. Surprisingly, this is true despite the high energy of these electronic excited states, a fact that has probably inhibited considering this mechanism before. The theory proposed in this work naturally explains the experimental correlation between the $g$-shift and relaxation time. Indeed, spin-orbit coupling is at the same time at the origin of the $g$-shift and a fundamental ingredient of $\hat{H}_{el}$ entering Eq. \ref{spinphonH}. An increased value of such interaction automatically leads to a stronger coupling between lattice and spin and to deviations to the free electron $g$-value. It is also important to remark that $\Gamma_I$ and $\Gamma_{II}$ mechanisms are operative at the same time. Depending on the relative strength of the crystal field and hyperfine coupling, either mechanism could become the leading one for a given molecule. To stress the importance of accounting for all the relaxation mechanisms in the interpretation of relaxometry experiments, we show in Fig. \ref{tauVO} the simulation of Raman relaxation time for the compound VO(dmit)$_2$\cite{atzori2016quantum} using both $\Gamma_I$ and $\Gamma_{II}$. Some of the present authors previously simulated the spin-lattice relaxation time for this molecule based on the $\Gamma_I$ mechanism and found excellent agreement with experiments\cite{lunghi2022toward}. The inclusion of $\Gamma_{II}$ contributions shows that the latter also accounts for similar relaxation rates, making it hard to attribute spin relaxation to a single mechanism, especially once numerical uncertainties are considered. 

Our study also shines new light on the identification of the phonons responsible for relaxation and highlights the limitations of using the common local-mode model for the interpretation of $T_1$ v.s. $T$ curves. The fitting of relaxation data often leads to the conclusion that modes with energy above 100 cm$^{-1}$ are necessary to explain spin-lattice relaxation. This interpretation is in disagreement with ab initio simulations of $T_1$, which instead points to low energy modes of tens of cm$^{-1}$ as the sole relevant. Our study confirms the latter interpretation, with the caveat that a larger window of low energy vibrations than previously noted should be removed to sensibly extend $T_1$. Since both ab initio methods and the phenomenological local-mode model descend from the same underlying assumptions\cite{lunghi2023spin}, e.g. weak coupling, Markovian dynamics, harmonic phonons, etc., the disagreement between the two suggests that conclusions based on the latter must be treated with considerable care and that the proposed higher energies for a local mode currently lacks an explanation. We also note that recent attempts to complement local-mode models with computed vibrational spectra have led to the reinforcement that high-energy modes are responsible for relaxation in spin-1/2 molecules\cite{kazmierczak2021impact,kazmierczak2024determining}. However, such attempts include adjustable coefficients and are invariably applied to gas-phase normal modes, which arguably differ from the real crystal vibrational density of states and provide a poor description of low-energy vibrations\cite{lunghi2019phonons}. Indeed, a very recent experimental study has shown that the modulation of the low-energy vibrational spectrum of vanadyl-based phthalocyanines deeply affects their $T_1$\cite{lohaus2024quantifying}, as predicted by previous simulations accounting for the full crystal vibrational spectrum \cite{garlatti2023critical}. At the same time, we acknowledge that although overall trends and relaxation rates are well reproduced, the calculated slope of $T_1$ v.s. $T$ deviates from the experimental one. Whilst discrepancies at low $T$ can be easily interpreted as the lack of acoustic phonons in the simulations, the high-$T$ behaviour cannot be easily reconciled with the current theoretical framework. The latter inevitably leads to the high-$T$ limit of $T_1\propto T^{-2}$, thus underestimating the experimental $T_1$ in the temperature range in which empirical fittings invoke a high-energy local mode. We perform numerical tests where we artificially increase the coupling to spin of high-energy phonons in the range individuated by the local-mode analysis and found that at least a factor four of increase in the coupling would be necessary to have a visible effect of the predicted $T_1$. Although a systematic investigation of the accuracy of electronic structure methods has not been performed yet, successful applications of our method to other compounds\cite{lunghi2022toward,mondal2022unraveling} make it improbable that numerical inaccuracies could lead to such a drastically different qualitative picture of relaxation. On the other hand, the accuracy of our methods in the prediction of low-energy vibrations in crystals has been thoroughly benchmarked against experiments\cite{garlatti2020unveiling,garlatti2023critical}. As such, we envision that the sophistication of the underlying open quantum systems methods is a priority for further improving the agreement between theory and experiments and conclusively assessing the role of high energy phonons in the spin relaxation of spin-1/2 systems. Interestingly, as this manuscript was being prepared, Shushkov reported a DFT-based computational method to predict the relaxation rate of  Cu(II) spin-1/2 molecules\cite{Shushkov2024novel}. Differently from the method presented here, Sushkov described an additional contribution to $\Gamma_I$ due to non-adiabatic effects. The latter contribution was found to dominate $T_1$ over the entire temperature range with respect to the modulation of \textbf{g} and to be driven by high-energy gas-phase molecular modes. Although this study also potentially suffers from a poor representation of low energy modes, the proposed relaxation mechanism stands out as an additional competing effect to those accounted for by $\Gamma_{II}$. To the best of our understanding, the same non-adiabatic effects would be fully included in $\Gamma_I$ by extending Eq. \ref{spinphonH} to the second order, making the possibility to compare all these methods within the same ab initio framework within reach\cite{mariano2023manual}.\\

Thanks to these new results on the interpretation of Raman relaxation, we are now in the position to critically review the list of chemical strategies at our disposal to control $T_1$. Indeed, achieving a large separation among ground and excited states as well as reducing the density of states in the low-energy part of the vibrational spectrum point to longer values of $T_1$. The different relaxation rates of CrN(pyrdtc)$_2$ and CrN(trop)$_2$ can indeed be interpreted following these considerations, with both lower excited states and lower energy phonons for the tropolonate. The overall remarkable properties of these square pyramidal complexes can also now be easily interpreted through the presence of the strongly $\sigma$ and $\pi$ donor nitrido group, which induces a very large splitting of $d$ orbitals . In light of these considerations, one would be tempted to employ heavier elements like 4$d$ or 5$d$ transition metals, which boast much larger crystal field splitting \cite{lever1968inorganic}. This however would also lead to an increased value of effective spin-orbit coupling, which is well documented as being detrimental for $T_1$\cite{chakarawet2021effect}. Interestingly, despite having provided a strong reinterpretation of Raman relaxation mechanism in spin-1/2 and the justification for its chemical control guidelines, the latter does not drastically shift from those previously inferred from considering the modulation of the $g$-tensor as the leading relaxation mechanism\cite{albino2019first,mirzoyan2020dynamic}. Incidentally, the $g$-shift and the residual orbital angular momenta of a molecule, often claimed to determine $T_1$\cite{albino2019first,ariciu2019engineering,mirzoyan2020dynamic}, also depend on the energy of excited states and have thus suggested large crystal field splitting as a design rule towards long-lived spin excitations. Similarly, for what concerns the phonons, both Raman mechanisms receive the largest contribution from low-energy vibrations, making previous advice still valid\cite{garlatti2023critical}, albeit with reduced expected effectiveness, as suggested by Fig. \ref{phon_convergence}.\\

In conclusion, our theoretical and experimental study has demonstrated that the interpretation of spin relaxation in spin-1/2 compounds requires going beyond considering only the ground state of molecules and including high-energy electronic virtual excitations. The inclusion of these contributions not only made it possible to reproduce experiments to an unprecedented level of accuracy but also provided novel conceptual tools to study and chemically engineer this important class of compounds. We envision that further development of ab initio open quantum system theory in synergy with novel experiments will soon make it possible to achieve a long-sought full reconciliation between theory and experiments.

\section*{Materials and Methods}

\textbf{Syntheses and samples preparation.} All solvents were HPLC grade from VWR chemicals, except for dichloromethane (DCM) which was from Fischer Chemical. Mn(N)Salen, and mer-Cr(Cl)$_3$(THF)$_3$ were synthesized as previously reported\cite{birk2003atom}. Tropolone (98\%) was obtained commercially from Combi Blocks, and recrystallized from isopropanol before use. CF$_3$COD (Sigma Aldrich), platinum on carbon, 10\% (Sigma Aldrich), and ammonium pyrrolidine dithiocarbamate (Acros) were used as received. CrN(pyrdtc)$_2$ and ReNCl$_2$(PPh$_3$)$_2$ were synthesized as described in the literature\cite{birk2003atom,DT9720001079}.

\textit{Synthesis of tropolone-d4:} The deuteration of tropolone followed literature protocols for aromatic compounds. Tropolone (1 g; 8.19 mmol) and 10\% Pt/C (0.2 g) were added to deuterium oxide (15 g; 75 mmol)\cite{sakakibara2021direct}. The reaction was kept acidic by addition of two drops of d-trifluoroacetic acid to subdue rearrangement of the tropolonate under the forcing conditions. The reaction was done in a Teflon lined steel autoclave at 145 C$^{\circ}$ for three days. Aqueous workup and recrystallization from isopropanol yielded 0.390 g (37.8\%) of Tropolone-d4. The deuteration degree was inferred to be 80\% from the HRMS of the resulting Cr(N)(trop-d4)$_2$ in agreement with the previously reported difference in reactivity of the 5-position compared to the remaining ring positions\cite{sakakibara2021direct}. 

\textit{Synthesis of Cr(N)(trop)$_2$ and Cr(N)(trop-d4)$_2$:} The procedure follows the general procedure described by one of the authors\cite{birk2003atom}. mer-Cr(Cl)$_3$(THF)$_3$ (0.153 g; 0.409 mmol) and Mn(N)(salen) (0.275 g; 0.821 mmol) were mixed in acetonitrile (10 mL) under a nitrogen atmosphere. The reaction was stirred at RT for 20 min. and the precipitated Mn(Cl)(salen) was filtered off. To the filtrate, tropolone (0.105 g; 0.859 mmol) in acetonitrile (3 mL) was added dropwise, while stirring under under N$_2$ upon formation of a brownish orange precipitate. After completion of the precipitate, the product was filtered off and washed with acetonitrile (2 times 30 mL) before being dried in a stream of N$_2$. The yield was 89.5 mg (71.1\%) of a brownish orange microcrystalline product. The polycrystalline sample was void of Cr(trop)$_3$ based on the comparison with the PXRD (Cu-K$\alpha$), of the latter. Crystals suitable for SC-XRD were obtained by recrystallization from acetonitrile/DCM (1:2). The deuterated version Cr(N)(trop-d4)$_2$ was synthesized by the same protocol yielding 70.5\% of isolated product with similar peak positions in the PXRD as the non-deuterated sample. HRMS, m/z=317.0604 (100\%), 316.0540 (92\%) corresponding to $>$ 80\% deuteration. 

\textit{Synthesis of ReN(pyrdtc)$_2$, Nitridobis (pyrrolidinedithiocarbamato)rhenium(V):} ReNCl$_2$(PPh$_3$)$_2$  (0.60 g, 0.75 mmol ) and NH$_4$(pyrdtc) (0.50 g, 3.0 mmol) were stirred in acetone (60 ml) overnight at 40$^\circ$C under N$_2$. The brownish solution containing a yellow precipitate was evaporated to dryness in a stream of N$_2$. The resulting solid was extracted with DCM/MeOH (9:1, 100 ml), filtered, and the extract was reduced in volume to ca. 10 ml yielding a bright yellow crystalline product, which was washed with ether, and dried in vacuum (0.24 g, 48\%). Crystals suitable for single crystal X-ray diffraction were obtained by slow evaporation of a DCM solution. 

\textit{X-ray structure determination:} Single crystals were obtained as described under the synthesis. They were coated with mineral oil, mounted on kapton loops, and transferred to the nitrogen cold stream of the diffractometer. Diffraction data were recorded at 100(2) K on a Bruker D8 VENTURE diffractometer equipped with a Mo K$\alpha$ high-brilliance I$\mu$S radiation source ($\lambda$ = 0.71073 \AA$ $), a multilayer X-ray mirror and a PHOTON 100 CMOS detector, and an Oxford Cryosystems low-temperature device. The instrument was controlled with the APEX3 software package using SAINT. Final cell constants were obtained from least squares fits of several thousand strong reflections. Intensity data were corrected for absorption using intensities of redundant reflections with the program SADABS. The structures were solved in Olex2 using SHELXT and refined using SHELXL. 

\textbf{Electron Paramagnetic Resonance Spectroscopy.} 
Solutions of 0.6 mM Cr(N)(pyrdtc)$_2$ and 0.5 mM CrN(trop)$_2$ were prepared by dissolution in CH$_2$Cl$_2$ followed by dilution with toluene to achieve a 3:7 CH$_2$Cl$_2$:toluene ratio. A solution of 0.5 mM CrN(trop-d4)$_2$ was prepared analogously in 3:7 CD$_2$Cl$_2$:toluene-d$_8$. Approximately 200 $\mu$L of solution or the doped $\sim$ 0.05\% Cr(N)(pyrdtc)$_2$ in ReN(pyrdtc)$_2$ solid was placed in 4 mm OD quartz tubes. Solutions were deoxygenated by multiple freeze-pump-thaw cycles.  All EPR samples were backfilled with $\sim$ 200 mtorr of helium prior to flame sealing the tube. The low-pressure He provides thermal contact between the sample and the walls of the tube to facilitate thermal equilibration. 

EPR spectroscopy and electron spin relaxation time measurements were performed on a Bruker E580 spectrometer with an Oxford ESR935 cryostat. X-band data were obtained with an ER4118X-MS5 split ring resonator and Q-band data were obtained with an ER5107 resonator. For pulse measurements the X-band resonator was over-coupled to a ringdown 1/e time of $\sim$ 2.5 – 3 ns which corresponds to $Q \sim 150$. The Q-band resonator was overcoupled to permit signal acquisition with less than 200 ns deadtime.  Temperature was controlled with a ColdEdge/Bruker Stinger closed-cycle helium cooling system. Temperature was monitored with a calibrated Lakeshore Cernox sensor and an Oxford Mercury iTC readout.  
Continuous wave (CW) spectra were measured at 60 or 70 K using low microwave power and 0.5 G modulation amplitude at 10 kHz to minimize passage effects. The spectra were simulated using the Bruker software Anisospinfit. 
$T_1$ was measured by 3-pulse inversion recovery with sequence $\pi - t - \pi/2 - \tau - \pi - \tau - echo $, with 2-step phase cycling, and the value of $t$ was incremented to record the recovery curve. The constant $\tau$ was set to 360 or 380 ns. The shot repetition time (SRT) was about 10 times the value of $T_1$ calculated with a stretched exponential. For the relatively high spin concentration in the 0.05\% CrN(pyrdtc)$_2$ in the Re analog the time constants were dependent on lengths of the $\pi/2$ pulse lengths between 16 and 160 ns. 
To minimize the impact of spectral diffusion on $T_1$ the inversion recovery data for 0.05\% CrN(pyrdtc)$_2$ were acquired with $\pi/2$ pulse lengths of 16 ns.  For 0.6 mM Cr(N)(pyrdtc)$_2$ in 3:7 CH$_2$Cl$_2$:toluene the values of $T_1$ were independent of pulse length, but for consistency with the doped solid the same pulse lengths were used. For 0.5 mM CrN(trop)$_2$ and CrN(trop-d4)$_2$ in 3:7 CH$_2$Cl$_2$:toluene or 3:7 CD$_2$Cl$_2$:toluene-d$_8$ $\pi/2$ pulse lengths of 40 ns were used. 
Inversion recovery curves were not single exponentials, indicating the presence of distributions of relaxation times. Inversion recovery data were fit with a stretched exponential 
\begin{equation}
    Y(t)=e^{-(t/T_1)^\beta}\:,
\end{equation}
and values of $\beta$ ranged from about 0.65 at 10 K to about 0.95 at 70 K. All fitted values are reported in the Supporting Information Tables S2 and S3.\\

\textbf{Electronic structure simulations.} Starting from X-ray structures of CrN(pyrdtc)$_2$ and CrN(trop)$_2$, cell and geometry optimization and simulations of $\Gamma$-point phonons have been performed with periodic density functional theory (pDFT) using the software CP2K\cite{kuhne2020cp2k}. Cell optimization was performed employing a very tight force convergence criteria of 10$^{-7}$ a.u. and SCF convergence criteria of 10$^{-10}$ a.u. for the energy. A plane wave cutoff of 1000 Ry, DZVP-MOLOPT Gaussian basis sets, and Goedecker-Tetter Hutter pseudopotentials were employed for all atoms. The  Perdew-Burke-Ernzerhof (PBE) functional\cite{perdew1996generalized} and DFT-D3 dispersion corrections\cite{grimme2010consistent} were used.

Electronic structure and magnetic properties were computed on the pDFT-optimized structures using the software ORCA\cite{neese2020orca}. The decontracted basis set DKH-def2-TZVPP, along with the Douglas-Kroll-Hess (DKH) scalar relativistic correction to the electronic Hamiltonian, was employed. Picture-change effects were included in the calculations. The computation of $\mathbf{g}$ and $\mathbf{A}$ tensors at the DFT level of theory was performed using a PBE0 exchange-correlation functional\cite{adamo1999toward} with the RIJCOSX approximation for Coulomb and Hartree-Fock exchange. Multireference calculations were performed using Complete Active Space Self Consistent Field (CASSCF) in combination with N-Electron Valence Perturbation Theory (NEVPT2). The active space used to build the CASSCF wavefunction is (1,5), i.e., one electron in five 3$d$-orbitals. All possible states with multiplicity two were included in the state-average procedure. Mean-field spin-orbit coupling operator, along with Quasi Degenerate Perturbation Theory (QDPT), was employed to account for the mixing of spin-free states.\\

\textbf{Spin-phonon relaxation simulations.} First-order vibronic coupling matrix elements of Eq. \ref{spinphonH} have been evaluated within the framework described before\cite{mariano2023manual}. Briefly, starting from the ab initio CASSCF wavefunctions $| \varphi_a \rangle$ and energies $E_a$, the matrix elements of the $\nabla \hat{H}_{el}$ operator are expressed using the Hellmann-Feynman theorem as
\begin{equation} \label{HF_theorem}
     \langle \varphi_a | \nabla \hat{H}_{el}| \varphi_b \rangle =\nabla E_b \delta_{ab} +(E_b -E_a) \langle \varphi_a|\nabla \varphi_b \rangle \:.
\end{equation}
To evaluate the non-adiabatic coupling (NAC) terms $\langle \varphi_a|\nabla \varphi_b \rangle$, numerical differentiation of wavefunction overlap has been performed using an in-house Python code interfaced with the program WFOVERLAP\cite{plasser2016efficient}. The step for the numerical differentiation has been set to 0.001 Å along the Cartesian coordinates of the system. A systematic convergence study of the $T_1$ relaxation time as a function of the differentiation step was performed. Once the vibronic coupling matrix elements have been computed in the Cartesian coordinates $\vec{R}$, they are transformed into the normal mode reference framework by using
\begin{equation}
     \left(\frac{\partial \hat{H}_\mathrm{el} }{\partial Q_{{\alpha}}} \right)= \sum_{a}^{3N} \sqrt{\frac{\hbar}{\omega_{\alpha}m_{a}}} L_{\alpha a}\left( \frac{\partial \hat{H}_\mathrm{el} } {\partial R_a} \right) \:,
    \label{cartesian_2_normal}
\end{equation}
where $m_i$ is the $i$-th atomic mass and $L_{\alpha i}$ the Hessian matrix eigenvectors. The latter is computed with a two-step numerical differentiation of forces. The vibronic coupling vectors from CASSCF, together with NEVPT2 energies in Eq. \ref{Red41_T}, are finally employed to simulate spin relaxation using Eq. \ref{Red41} with the software MolForge\cite{lunghi2022toward}. The use of CASSCF vibronic coupling, in conjunction with NEVPT2 vertical excitation energies, is necessary here, as orbital relaxation at the NEVPT2 level is not supported in ORCA. The Dirac delta functions appearing in the spin dynamics equations are broadened with a Gaussian smearing of 30 cm$^{-1}$ after a convergence study on the computed $T_1$.  

The evaluation of the second-order terms of Eq. \ref{sph} requires computing the second-order derivatives of the tensors $\mathbf{A}$ and $\mathbf{g}$, namely $(\partial^2 \mathbf{A}/\partial q_\alpha \partial q_\beta)$ and $(\partial^2 \mathbf{g}/\partial q_\alpha \partial q_\beta)$. In this study, we employed the machine-learning-based strategy to calculate the second-order derivatives of the spin Hamiltonian in Eq. \ref{spinH}\cite{lunghi2020limit,garlatti2023critical}. These coefficients are subsequently employed to simulate the time-dependency of the $z$-component of the simulated magnetization $\vec{\mathbf{M}}$. Finally, $T_1$ is extracted by fitting $M_z$ with a double exponential function.

Calculations of relaxation rates and magnetization time evolution were obtained with the development branch \textit{T4} of the software MolForge\cite{lunghi2022toward}, available at
\textit{github.com/LunghiGroup/MolForge}.

\vspace{0.2cm}
\noindent
\textbf{Acknowledgements and Funding}\\
This project has received funding from the European Research Council (ERC) under the European Union’s Horizon 2020 research and innovation programme (grant agreement No. [948493]). Computational resources were provided by the Trinity College Research IT and the Irish Centre for High-End Computing (ICHEC).

\vspace{0.2cm}
\noindent
\textbf{Authors Contributions}\\
A.L., S.S.E., and G.R.E. conceived the project. L.M. and A.L. developed the theoretical and computational models. A.L., L.M., and V.H.A.N carried out the calculations and their analysis. G.R.E. and S.S.E. performed the EPR measurements. J.B., J.B.P. and M.B. synthesized the compounds. All authors contributed to the discussion of results, and to writing the manuscript.

\vspace{0.2cm}
\noindent
\textbf{Conflict of interests}
The authors declare no competing interests.


\end{document}